\pdfoutput=1
\documentclass[a4paper]{article}

\usepackage{INTERSPEECH2020}
\usepackage{xcolor}
\usepackage{cite}

\title{%
Multi-talker ASR for an unknown number of sources:\\
Joint training of source counting, separation and ASR%
}
\name{Thilo von Neumann$^{1,2}$, Christoph Boeddeker$^1$, Lukas Drude$^1$, Keisuke Kinoshita$^2$, Marc Delcroix$^2$, Tomohiro Nakatani$^2$, Reinhold Haeb-Umbach$^1$}
\address{
    $^{1}$Paderborn University, Department of Communications Engineering, Paderborn, Germany \\
    $^{2}$NTT Corporation, Japan
}

\email{%
\{vonneumann, boeddeker, drude, haeb\}@nt.upb.de\\\{keisuke.kinoshita, marc.delcroix, tnak\}@ieee.org%
}

\usepackage{amsmath}
\usepackage{amssymb}
\usepackage{bbm}
\usepackage{physics}
\usepackage{siunitx}
\usepackage{trfsigns}
\usepackage{pifont}
\usepackage{etoolbox}
\usepackage{multirow}
\usepackage{csquotes}
\usepackage[shortlabels]{enumitem}  % Allows \begin{enumerate}[label=\Alph*] ... \end{enumerate}
\usepackage[capitalize]{cleveref}
\usepackage{tikz}
\usepackage{balance} % For the balanced reference page

\newcolumntype{H}{>{\setbox0=\hbox\bgroup}c<{\egroup}@{}}   % Hidden column
\newcommand{\mrow}[1]{\multirow{2}[2]{*}{\begin{tabular}{@{}c@{}}#1\end{tabular}}}  % Centered row if using \cmidrule
\newcommand{\tbli}{\phantom{+}}
\newcommand{\tblii}{\phantom{++}}

\newcommand*{\thl}{\fontseries{b}\selectfont}
\robustify\thl
\robustify\fontseries
\newcommand{\loss}[1]{\mathcal{L}^{\text{(#1)}}}

\usepackage[acronym]{glossaries}
\newacronym{ASR}{ASR}{automatic speech recognition}
\newacronym{DPRNN}{DPRNN}{Dual-Path Recurrent Neural Network}
\newacronym{TasNet}{TasNet}{Time-domain audio separation Network}
\newacronym{OR-PIT}{OR-PIT}{One-and-Rest Permutation Invariant Training}
\newacronym{CTC}{CTC}{Connectionist Temporal Classification}
\newacronym{WER}{WER}{Word Error Rate}
\newacronym{CER}{CER}{Character Error Rate}
\newacronym{PIT}{PIT}{Permutation Invariant Training}
\newacronym{DPCL}{DPCL}{Deep Clustering}
\newacronym{BLSTM}{BLSTM}{Bi-directional Long Short-Term Memory Network}

\begin{document}

\maketitle
\begin{abstract}
% max of 200 words
% 
% Problem: multi-speaker speech recognition with source number counting
% Why: There is a lot of overlap in common scenarios and the number of speakers is typically not known
% How: Joint training of an iterative source extraction scheme with ASR
% Results: Very good counting accuracy, Improvement in WER over previous works
% 
Most approaches to multi-talker overlapped speech separation and recognition assume that the number of simultaneously active speakers is given, but in realistic situations, it is typically unknown. 
To cope with this, we extend an iterative speech extraction system with mechanisms to count the number of sources and combine it with a single-talker speech recognizer to form the first end-to-end multi-talker automatic speech recognition system for an unknown number of active speakers.
Our experiments show very promising performance in counting accuracy, source separation and speech recognition on simulated clean mixtures from WSJ0-2mix and WSJ0-3mix. 
Among others, we set a new state-of-the-art word error rate on the WSJ0-2mix database.
Furthermore, our system generalizes well to a larger number of speakers than it ever saw during training, as shown in experiments with the WSJ0-4mix database.

%Due to the large amount of overlapped speech in recordings of common scenarios, such as meetings, interest raised in the task of multi-talker speech recognition. 
%Different types of multi-talker speech recognizers have been proposed with promising performance.
%Many of these, however, assume that the number of talkers present is given, but in realistic situations it is typically unknown.
%We propose to integrate a source number counting technique into a multi-talker speech recognizer.
%We extend an iterative speech extraction system with mechanisms to count the number of talkers and combine it with a single-talker speech recognition to form an end-to-end multi-talker automatic speech recognizer for unknown numbers of talkers.
%Our experiments show very promising performance in counting accuracy, source separation and speech recognition on simulated clean mixtures from WSJ0-2mix and WSJ0-3mix.
%Our system generalizes to a larger number of talkers that it never saw during training, i.e., four-talker mixtures from the WSJ0-4mix database.

\end{abstract}

\section{Introduction}

Overlapping speech is quite common for many scenarios, as the meetings recorded in the AMI corpus \cite{McCowan2005_AMIMeetingCorpus} show overlap in about \SIrange{5}{10}{\%} of the time, and in informal situations, as recorded in the CHiME-5 \cite{Barker2018_FifthCHiMESpeech} challenge data, it can exceed \SI{20}{\%}.
A typical task for analysis of such recordings is \gls{ASR}.
In the case of overlapping speech, this is called multi-talker speech recognition where the speech of multiple concurrently active talkers is to be recognized.
Conventional speech recognizers are limited to handling a single talker at a time which makes them inapplicable in those scenarios.

Many efforts have already been put into the field of multi-talker speech recognition \cite{Seki2018_PurelyEndtoEndSystem,Chang2019_EndtoendMonauralMultispeaker,Kanda2020_SerializedOutputTraining,Settle2018_EndtoendMultispeakerSpeech,Menne2019_AnalysisDeepClustering,vonNeumann2020_EndtoEndTrainingTime,Yu2017_RecognizingMultiTalkerSpeech,Qian2018_SinglechannelMultitalkerSpeech,Bahmaninezhad2019_ComprehensiveStudySpeecha}.
The approaches can basically be divided into monolithic and cascade systems.
The monolithic ones form one large neural network that is optimized as a whole, e.g., by extending a single-talker CTC/attention ASR system so that its encoder outputs one latent representation for each speaker in the mixture and one or multiple attention decoders reconstruct one transcription for each representation \cite{Seki2018_PurelyEndtoEndSystem,Chang2019_EndtoendMonauralMultispeaker}.
% These systems form one large neural network, so they are end-to-end monolithic models
Another monolithic approach called Serialized Output Training modifies the target label sequence of a single-speaker recognizer to output the transcriptions for all talkers serially delimited by a special \enquote{speaker change} token \cite{Kanda2020_SerializedOutputTraining}.
These systems have the disadvantage that they do not provide interpretable signals as, e.g., separated speech signals.

\begin{figure}[t]
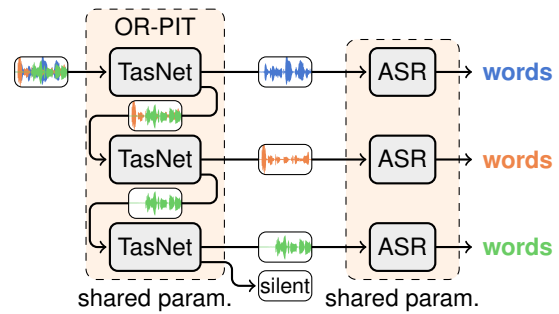

    \centering
    \include{tikz/architecture}
    \vspace{-\baselineskip}
    \caption{Example of the proposed iterative multi-speaker speech recognition system processing a three-speaker mixture.}
    \label{fig:architecture}
    % \vspace{-\baselineskip}
\end{figure}

The cascade systems use source separation techniques followed by single-talker speech recognizers.
These systems have the advantage that intermediate signals are interpretable and individual system parts can be trained and tested on their own.
As separation front-ends, \gls{DPCL} \cite{Isik2016_SingleChannelMultiSpeakerSeparation}, \gls{PIT} and the currently most promising \gls{TasNet} architecture \cite{Luo2018_TaSNetTimeDomainAudio} have been combined with different speech recognizers \cite{Settle2018_EndtoendMultispeakerSpeech,Menne2019_AnalysisDeepClustering,Yu2017_RecognizingMultiTalkerSpeech,Isik2016_SingleChannelMultiSpeakerSeparation,Bahmaninezhad2019_ComprehensiveStudySpeecha,vonNeumann2020_EndtoEndTrainingTime}.
The experiments in general suggest that having both, dedicated parts for the separation and the recognition part, and joint end-to-end fine-tuning is beneficial.

Most separation and multi-talker ASR approaches assume that the number of talkers is known \cite{Kolbaek2017_MultitalkerSpeechSeparationa,Luo2018_TaSNetTimeDomainAudio,Luo2020_DualPathRNNEfficient,Seki2018_PurelyEndtoEndSystem,Chang2019_EndtoendMonauralMultispeaker,Settle2018_EndtoendMultispeakerSpeech,Qian2018_SinglechannelMultitalkerSpeech}, although this is not the case in realistic situations.
Source number counting has been combined with source separation in iterative speech extraction  \cite{Kinoshita2018_ListeningEachSpeaker,Takahashi2019_RecursiveSpeechSeparation} and model selection schemes \cite{Nachmani2020_VoiceSeparationUnknown}, but not yet with speech recognition.

We propose the first jointly optimized multi-talker ASR system for an unknown number of speakers by combining source separation and number counting techniques from \cite{Luo2020_DualPathRNNEfficient,Kinoshita2018_ListeningEachSpeaker,Takahashi2019_RecursiveSpeechSeparation} with a single-speaker CTC/attention speech recognizer \cite{Watanabe2017_HybridCTCAttention,Kim2017_JointCTCattentionBased,Watanabe2018_EspnetEndtoendSpeech,Xiao2018_HybridCTCattentionBased}.
To do that, we first investigate a \gls{DPRNN}-TasNet separator for fixed numbers of speakers as a front-end for ASR.
We jointly fine-tune the \gls{DPRNN} with a pre-trained ASR system, which was shown to be effective in other scenarios \cite{Heymann2017_BeamnetEndtoendTraining} and multi-talker ASR \cite{vonNeumann2020_EndtoEndTrainingTime}, to optimize the overall model performance.
By doing so, we achieve new state-of-the-art performance of \SI{7.5}{\%} \gls{WER}.
We then integrate the \gls{DPRNN} into the \gls{OR-PIT} architecture and extend it with elegant mechanisms for source number counting.
The counting mechanisms show a promising performance.
Finally, we combine the \gls{OR-PIT} with an ASR system to form our new multi-speaker ASR system for unknown numbers of speakers.
Experiments show that our system generalizes to a larger number of speakers than it saw during training.

\section{Source separation and counting}
In the following two subsections we first introduce the baseline approach to source separation, where the maximum number of talkers is assumed to be known.
Then, we generalize it to separate an a priori unknown number of concurrent talkers.

% \subsection{Source Separation with the Dual-Path RNN TasNet}
% \CB{alternative section title suggestion:}
\subsection{Known number of talkers: Dual-Path RNN-TasNet}
\def\mix{x(t)}
\def\sourceletter{s}
\def\source{\sourceletter(t)}
\def\spkidx{k}
\def\sourceindexed#1{\sourceletter_{#1}(t)}
\def\sourcehat{\hat{\sourceletter}(t)}
\def\sourcek{\sourceletter_\spkidx(t)}
\def\sourcekhat{\hat{\sourceletter}_\spkidx(t)}
\def\spkcount{K}
\def\tasnetsegmentlength{L}
\def\sisnrscalingfactor{\alpha}

A single channel discrete-time speech mixture signal $\mix$ is modeled as a sum of $K$ single-talker speech signals $\sourcek$:
\begin{equation}
    \mix = \sum_{\spkidx = 1}^{\spkcount} \sourcek,
\end{equation}
where $\spkidx$ is the talker index.
Source separation is concerned with extracting the source signals $\sourcek$ from the mixture $\mix$.

We use the \gls{TasNet} \cite{Luo2018_TaSNetTimeDomainAudio,Luo2019_ConvTasNetSurpassingIdeal} with a \gls{DPRNN} \cite{Luo2020_DualPathRNNEfficient} as the separation network to obtain a fixed number of estimations $z_k(t)$ for the sources $\sourcek$, where the order of $z_k(t)$ can be permuted to the actual source signals:
\begin{equation}
    [z_1 (t), ..., z_K (t)] = \text{DPRNN-TasNet}\big(\mix\big).
\end{equation}
It works by encoding the time-domain signal into a latent domain with a convolutional encoder, separating this representation with the \gls{DPRNN}, and transforming it back to time domain with a de-convolutional decoder.
The \gls{DPRNN} models short- and long-term dependencies in an alternating manner by segmenting the input and skipping different numbers of time-steps in adjacent layers.
It was shown in \cite{Luo2020_DualPathRNNEfficient} to be superior to a separator based on 1-D convolutions used in the Conv-\gls{TasNet} \cite{Luo2019_ConvTasNetSurpassingIdeal}.

% ... I don't know yet how detailed this can be ...

% \begin{itemize}
%     \item Description of the DPRNN-TasNet architecture
%     \begin{itemize}
%         \item Encoder: 1-D convolution as in TasNet but with a bias term
%         \begin{equation}
%             \vect{w} = \mathcal{H}(\vect{x}\matr{U} + \vect{b})
%         \end{equation}
%         for signal segments $\vect{x}$. Concatenate the resulting vectors to form the encoded feature sequence $\matr{W} = [\vect{w}_i]$
%         \item Segmentation: Cut the encoded feature sequence $\matr{W}$ into segments with \SI{50}{\%} overlap and concatenate them to a three-dimensional tensor
%         \item Process alternatingly along time and segment (chunk?) direction to model short- and long-term dependencies alternatingly
%         \item Overlap-add to obtain the final encoded feature sequence
%         \item Decoder: 1-D transposed convolution as in TasNet
%         \begin{equation}
%             \hat{\vect{x}} = \vect{w}\matr{V} + \vect{b}
%         \end{equation}
%     \end{itemize}
% \end{itemize}

\subsubsection{Time-domain training objective}

In recent work, the training objective often is to maximize the scale-invariant source-to-distortion ratio (SI-SDR)\footnote{Sometimes called SI-SNR \cite{Luo2018_TaSNetTimeDomainAudio,Luo2019_ConvTasNetSurpassingIdeal}} \cite{Luo2018_TaSNetTimeDomainAudio,Luo2019_ConvTasNetSurpassingIdeal} by minimizing the negative SI-SDR.
%
% \begin{align}
%     \loss{SI-SDR} \big(\source, \sourcehat\big) &= -10 \log_{10} \frac{\sum_t |\sisnrscalingfactor \source|^2}{\sum_t |\sisnrscalingfactor \source - \sourcehat |^2},\\
%     \text{with }\sisnrscalingfactor &= \frac{\sum_t \sourcehat \source}{\sum_t |\source|^2}.
% \end{align}
%
% 
%
% While this formulation makes the loss scale-invariant,
% , i.e., a constant factor in the estimated signals has no impact on the loss value and gradients
% it does not allow silence as a source signal, i.e., $\source = 0$.
% \inred{Mention gradient issues?}
By setting $\alpha=1$ and removing terms that do not depend on $\sourcehat$, the time-domain logarithmic MSE loss can be obtained \cite{Heitkaemper2020_DemystifyingTasNetDissecting}:
% that allows completely silent target signals at the cost of not being scale-invariant anymore:
%
\begin{equation}
    \loss{T-LMSE}\big(\source, \sourcehat\big) = 10 \log_{10} \sum_t |\source - \sourcehat|^2.
\end{equation}
The missing scale-invariance did not show negative effects on the separation performance \cite{Heitkaemper2020_DemystifyingTasNetDissecting}.
To be able to handle different numbers of speakers with such a model, it is common for frequency-domain separators to set the missing outputs to silence \cite{Kolbaek2017_MultitalkerSpeechSeparationa}, i.e., $\source=0$.
To use silent targets here, the loss has to be prevented from going to negative infinity for perfect reconstruction and masking any loss terms from other target signals by, e.g., adding $1$ to the argument of the logarithm:
%
% Training on silent targets, i.e., $x(t) = 0$ is still problematic as the loss for perfect reconstruction goes to negative infinity $\loss{T-LMSE} (0, 0) \rightarrow -\infty$, which shadows any other loss terms from other targets that aim to reconstruct speech.
% To handle this, a constant value can be added to the operand of the logarithm to prevent it from becoming $0$ while keeping the logarithmic properties:
%
\begin{equation}
    \loss{T-L1PMSE}\big(\source, \sourcehat\big) = 10 \log_{10}\big(1 + \sum_t |\source - \sourcehat|^2\big).
\end{equation}
The total loss is calculated in a permutation-invariant manner with the set $\mathcal{P}$ of all permutations of length $K$:
\begin{equation}
    \loss{reconstruction} = \min_{\phi\in\mathcal{P}} \frac{1}{\spkcount} \sum_{k=1}^{K}\mathcal{L}(\sourceindexed{\phi(k)}, z_k (t)).
\end{equation}

\def\na{{n/a}}

\subsection{Unknown number of talkers: OR-PIT}

Conventional source separators are limited to a fixed number of talkers.
An arbitrary number of talkers can be handled by iterative source extraction approaches \cite{Kinoshita2018_ListeningEachSpeaker,Takahashi2019_RecursiveSpeechSeparation}.
Instead of directly separating the mixture into one stream for each talker, they apply a network repeatedly to extract one talker at a time.

\def\orpitprimout{z_1(t)}
\def\orpitsecout{z_2(t)}
Following the \gls{OR-PIT} \cite{Takahashi2019_RecursiveSpeechSeparation} scheme, an iterative source extractor is a two-output separator, in our case a DPRNN-TasNet, trained to output one talker at its primary output $\orpitprimout$ and the sum of all remaining talkers at its secondary output $\orpitsecout$ so that $\orpitsecout$ can be fed back to extract the next talker, as visualized in the left part of \cref{fig:architecture}.
It is first trained with
\begin{equation}
  \label{eq:or-pit}
    \loss{OR-PIT} = \min_\spkidx \mathcal{L} \big(\orpitprimout, \sourcek\big) + \mathcal{L}\big(\orpitsecout, \sum_{i\neq\spkidx} \sourceindexed{i}\big)
\end{equation}
on clean mixtures, where $\mathcal{L}$ is a time-domain loss.
It is then fine-tuned by feeding $\orpitsecout$ as additional training data.

The number of talkers can be determined by counting the number of iterations required, until $\orpitsecout$ does not contain speech.
The authors of \cite{Takahashi2019_RecursiveSpeechSeparation} use Alexnet \cite{Sutskever2012_ImagenetClassificationDeep} as an external classifier to detect the absence of speech.
A stopping criterion can, however, be integrated into the separator by using thresholding or an additional output flag, as inspired by \cite{Kinoshita2018_ListeningEachSpeaker}.

% The absence of speech can be detected in various ways, discussed in the following paragraphs.
%
% \begin{enumerate}[(a)]
%     \item an energy-based threshold,
%     \item an external classifier, or
%     \item an additional output flag.
% \end{enumerate}
%
% \subsubsection{Energy-based threshold}

Assuming that a speech signal contains a certain minimal amount of average energy and the network is able to suppress speech well enough, absence of speech can be detected by:
%checking the average energy in the secondary output:
% The secondary output is used in this work. However, it is possible to use the estimated signal as well and this actually worked better in the experiments in my thesis, but I don't have joint models trained for this case.
%
% \def\threshold{\gamma^{\text{(threshold)}}}
\def\threshold{\gamma}
\begin{equation}
    \frac{1}{T} \sum_t |\orpitsecout|^2 < \threshold,
\end{equation}
with $T$ being the length of the signal and the threshold value $\threshold$ being determined manually.
Models that use this stopping criterion are called \enquote{threshold} models.
% \CB{Please verify, that this is the implementation. This thresholding is unreasonable, when SI-SDR is used.}

\def\estflag{\hat{f}}
\def\targetflag{f}
An elegant way to integrate the classifier into the separator is to let the separation network predict a stop flag.
This is done by adding an additional scalar output $\targetflag$ to the network.
% by increasing the output dimension of the DPRNN, mapping to a scalar for every time-step with a fully connected layer and averaging over time to obtain a scalar.
The output size of the \gls{DPRNN} is increased. 
The newly added \gls{DPRNN} outputs are transformed by a fully connected layer to a scalar for each time step.
The stop flag per utterance is obtained with, e.g., an average pooling over time followed by a sigmoidal function.
The estimated flag $\estflag$ is trained to indicate when to stop iterating using a binary cross-entropy objective
\begin{equation}
    \loss{flag} = - \targetflag \log(\estflag) - (1 - \targetflag) \log (1 - \estflag).
\end{equation}
%
% The target flag $\targetflag$ is set to $0$ while there is speech left to be extracted, and to $1$ when the separation is finished.
The target flag $\targetflag$ is set to $1$ when the second output should be empty and $0$ otherwise.

% \section{CTC/attention speech recognition}

% The CTC/attention speech recognizer combines \gls{CTC} with attention mechanisms from an end-to-end speech recognizer.
% The architecture we used is similar to \cite{Watanabe2017_HybridCTCAttention} and taken from the ESPnet toolkit \cite{Watanabe2018_EspnetEndtoendSpeech}.
% We replaced the original feature extraction with differentiable STFT-based features.
% It is trained with a multi-target loss function
% %
% \begin{equation}
%     \loss{ASR} = \lambda \loss{CTC} + (1 - \lambda)\loss{att},
% \end{equation}
% %
% with the manually set factor $\lambda=0.2$.
% \CB{with $\lambda = ???$.}

% The overall performance of the system could be improved by using a stronger speech recognizer as, e.g., the transformer model \cite{Dong2018_SpeechTransformerNoRecurrenceSequencetoSequence}, but the general observed tendencies remain the same.

\section{Joint optimization of source separation and speech recognition}

We propose to jointly fine-tune a source separation front-end (FE) with source number counting and a speech recognition back-end.
The FE is either a TasNet or an OR-PIT and the back-end is a CTC/attention speech recognizer similar to \cite{Watanabe2017_HybridCTCAttention,Seki2018_PurelyEndtoEndSystem} taken from the ESPnet toolkit \cite{Watanabe2018_EspnetEndtoendSpeech}.
We replace the original feature extraction with differentiable STFT-based features.
It is trained with a multi-target loss function with the factor $\lambda=0.2$:
\begin{equation}
    \loss{ASR} = \lambda \loss{CTC} + (1 - \lambda)\loss{att}.
\end{equation}

Joint fine-tuning is straight forward for the TasNet-based system without counting.
The gradients are propagated from the back-end into the FE and their losses are combined like
\begin{equation}
    \mathcal{L} = \loss{ASR} + \loss{FE}.
\end{equation}
%
% The model is not sensitive to weighting of the losses.
We solve the permutation problem with the FE signal-level loss.
% \CB{How important is the FE loss?}
% Different parts of the model can be fine-tuned exclusively by only updating their model weights.

For the iterative OR-PIT system, there are different options that arise from the fact that $\orpitsecout$ can contain more than one talker and is, thus, unusable for training the back-end.
We formulate two training schemes, namely the single- and multi-iteration schemes.
The single-iteration scheme unrolls a single iteration of the OR-PIT FE and uses $\orpitprimout$ to train the back-end.
$\orpitsecout$ is optimized using only the signal-level loss.
Here, the OR-PIT always sees unprocessed data, so there is a mismatch to evaluation where its own output is fed back.
% Other than that, this halves the training data available for the back-end and the loss functions used for the primary and secondary outputs are inconsistent.
To mitigate this, the model can be unrolled in the multi-iteration scheme, where the secondary output is used as the input for the following iteration and all primary outputs are used to train the ASR back-end.

% Note that these systems become quite large in terms of memory consumption.
% Previous works used some tricks to fit the model on a single GPU \cite{vonNeumann2019_EndtoendTrainingTime}, but our DPRNN implementation is small enough to fit together with the joint model on a single GPU without such tricks.

\section{Experiments}

We evaluate our systems in terms of average improvement of signal-to-distortion ratio (SDRi)\footnote{Provided by the mir\_eval toolbox \cite{Raffel2014_MirEvalTransparent}.} \cite{Vincent2006_PerformanceMeasurementBlind,Fevotte2005_BSSEVALToolbox}, \gls{CER}, \gls{WER} and counting accuracy on the WSJ, WSJ0-2mix, WSJ0-3mix and WSJ0-4mix databases \cite{Paul1992_DesignWallStreet,Isik2016_SingleChannelMultiSpeakerSeparation,Kolbaek2017_MultitalkerSpeechSeparationa} with a sample rate of \SI{8}{kHz}.
We use WSJ0-4mix as created in \cite{Takahashi2019_RecursiveSpeechSeparation}.
The experiments on source separation and number counting are conducted on the \emph{min} subsets of the WSJ0-mix databases that contain full overlap only.
Speech recognition is evaluated on the \emph{max} subset where no utterances are truncated.

\subsection{Source Separation}
\begin{table}[t]
    \centering
    \caption{
    Source separation performance in SDRi in dB for varying numbers of talkers given the oracle number of sources. The mixtures contain full overlap only (min subset).
    }\label{tab:separation}
    \setlength{\tabcolsep}{3pt}
    \begin{tabular}{ll SSSS}
        \toprule
        \multicolumn{2}{l}{\mrow{Method}} & \multicolumn{4}{c}{number of talkers} \\
        \cmidrule{3-6}
        && {1} & {2} & {3} & {4} \\
        \midrule
        
        \multirow{5}{*}{\rotatebox[origin=c]{90}{DPRNN-}}&TasNet 2 talker & 10.8 & 17.0 & {---} & {---} \\
        &TasNet 3 talker & 8.2 & 11.7 & 11.3 & {---} \\
        &TasNet 2+3 talker & \thl 39.4 & 17.0 & 10.9 & {---} \\
        % DPRNN 1+2+3 spk & {---} & {---} && {---}  & {---} && {---} & {---} && \na & \na \\
        \cmidrule{2-6}
        &OR-PIT stop-flag & 30.6 & 17.2 & 12.5 & 8.7 \\
        &OR-PIT threshold & 30.9 & \thl 17.5 & 12.8 & 8.4 \\
        \midrule
        \multicolumn{2}{l}{RSAN stop-flag \cite{Kinoshita2018_ListeningEachSpeaker}} & {---} & 8.6 & {---} & {---}\\
        % \cite{Nachmani2020_VoiceSeparationUnknown} & {---} & {---} && {20.1}  & {0.37} && {16.9} & {0.41} && {12.9} & {0.47} \\
        \multicolumn{2}{l}{Conv-TasNet-OR-PIT \cite{Takahashi2019_RecursiveSpeechSeparation}} & {---} & 15.0 & \thl 12.9 & \thl 10.6 \\
        \multicolumn{2}{l}{Original DPRNN-TasNet \cite{Luo2020_DualPathRNNEfficient}} & {---} & 16.1 & {---} & {---} \\
        \bottomrule
    \end{tabular}
    % \vspace{-\baselineskip}
\end{table}
\noindent
We first assume that the number of talkers is known.
We train all source separators on \SI{4}{s} long signal segments to comply with \cite{Luo2018_TaSNetTimeDomainAudio,Luo2020_DualPathRNNEfficient}.
We choose the DPRNN parameters according to the original paper \cite{Luo2020_DualPathRNNEfficient}%
%\footnote{\CB{Publish the code to padertorch?}}
, i.e., six blocks with two \glspl{BLSTM} with 128 units, an encoder window size of 16 and a chunk size of 100.
We train two \gls{DPRNN}-\glspl{TasNet} for two or three speakers only, respectively, and one with three outputs for two and three speakers where the last target is set to silence for two-speaker mixtures.
Our OR-PIT is trained on single-, two- and three-speaker recordings.
We use $\loss{T-L1PMSE}$ for \enquote{DPRNN-TasNet 2+3 talker} to cope with silent output, and $\loss{T-LMSE}$ for all other models.

The separation performance displayed in \cref{tab:separation} is evaluated given the oracle number of talkers, so that counting issues are not reflected in the SDRi.
The OR-PIT is forced to the correct number of iterations.
For the TasNet, only the $K$ outputs with the largest energy are considered for evaluation.

Our two-talker TasNet (DPRNN-TasNet 2 talker) achieves a separation performance slightly better than \cite{Luo2020_DualPathRNNEfficient} on two talkers and the same architecture works well on three talkers (DPRNN-TasNet 3 talker).
Both do not generalize well to smaller numbers of talkers.
The models specialized to one specific number are not able to produce a better reconstruction quality than what they learned during training.
The model trained on both, two and three speakers, handles both types of mixtures well and generalizes even to single-speaker recordings.

The OR-PIT can handle one to three speakers slightly better than the TasNet.
It especially has advantages in the three-speaker case, as also observed in \cite{Takahashi2019_RecursiveSpeechSeparation}, where the OR-PIT network extracts one speaker at a time while the TasNet has to solve the more complex task of separating three speakers simultaneously.
Its iterative structure allows the model to generalize to some extent to a larger number of talkers.
The performance, however, drops compared to the original Conv-TasNet-OR-PIT.
This might be caused by slight differences in training scheme.

% TODO: OR-PIT seems to perform better on larger numbers of speakers compared to TasNet (compare OR-PIT paper)

% \begin{table*}[h]
% \centering
% \caption{Source separation performance of the DPRNN-OR-PIT model given the oracle number of sources.}
% \begin{tabular}{ccSScSScSS}
%     \toprule
%     \mrow{Fine-tuned}&\mrow{Flag}& \multicolumn{2}{c}{\tab{1 Speaker\\(WSJ)}} && \multicolumn{2}{c}{\tab{2 Speakers\\(WSJ0-2mix)}} && \multicolumn{2}{c}{\tab{3 Speakers\\(WSJ0-3mix)}} \\
%     \cmidrule{3-4}\cmidrule{6-7}\cmidrule{9-10}
%     & & {SI-SDRi} & {SDRi} && {SI-SDRi} & {SDRi}  && {SI-SDRi} & {SDRi} \\
%     \midrule
%     \xmark&\xmark & 69.4 & 61.1 && 16.3 & 16.6 && 10.9 & 11.4 \\
%     \xmark&\cmark & 73.7 & 62.0 && 15.7 & 16.1 && 10.1 & 10.7\\
%     \cmark&\xmark & 30.8 & 30.9 && 17.2 & 17.5 && 12.3 & 12.8 \\
%     \cmark&\cmark & 30.5 & 30.6 && 16.9 & 17.2 && 12.0 & 12.5 \\
%     \midrule
%     \multicolumn{2}{l}{2-speaker DPRNN} & {---} & {---} && 16.7 & 17.0 && {---} & {---} \\
%     \multicolumn{2}{l}{Conv-TasNet OR-PIT \cite{Takahashi2019_RecursiveSpeechSeparation}} & {---} & {---} && 14.8 & 15.0 && 12.6 & 12.9 \\
%     \multicolumn{2}{l}{Oracle Binary Mask \cite{Takahashi2019_RecursiveSpeechSeparation}} & {$+\infty$} & {$+\infty$} && 13.0 & 13.5 && 13.2 & 13.6 \\
%     \bottomrule
% \end{tabular}
% \end{table*}

\subsection{Source Number Counting}

\begin{table}[t]
    \centering
    \caption{
    Source number counting accuracy in \% for varying numbers of talkers. The mixtures contain full overlap only (min subset). ($^*$: Threshold not optimized for this number of sources)
    }\label{tab:counting}
    \setlength{\tabcolsep}{2pt}
    \begin{tabular}{ll SSSS}
        \toprule
        \multicolumn{2}{l}{\mrow{Method}} & \multicolumn{4}{c}{number of talkers} \\
        \cmidrule{3-6}
         && {1} & {2} & {3} & {4} \\
        \midrule
        \multirow{3}{*}{\rotatebox[origin=c]{90}{DPRNN-}}
        & TasNet 2+3 talker & 0.0{$^*$} & 99.9 & \thl 99.4 & {---} \\
        % DPRNN 1+2+3 spk & {---} & {---} && {---}  & {---} && {---} & {---} && \na & \na \\
        \cmidrule{2-6}
        & OR-PIT stop-flag & \thl 100.0 & \thl 100.0 & 96.7 & \thl 91.9 \\
        & OR-PIT threshold & \thl 100.0 & \thl 100.0 & 98.2 & 0.0{$^*$}\\
        \midrule
        \multicolumn{2}{l}{RSAN stop-flag \cite{Kinoshita2018_ListeningEachSpeaker}} & \thl 100.0 & 99.9 & {---} & {---} \\
        \multicolumn{2}{l}{Model selection \cite{Nachmani2020_VoiceSeparationUnknown}} & {---} & 37.0 & 41.0 & 47.0 \\
        \bottomrule
    \end{tabular}
    % \vspace{-\baselineskip}
\end{table}

\begin{table*}[ht]
\centering
\caption{
Recognition performance of the multi-talker ASR systems for varying numbers of speakers given the oracle number of sources. The mixture signals contain full utterances (max subset). (\enquote{FE}: front-end, \enquote{ASR}: speech recognition part.)}
% \footnotesize
\label{tab:or-pit-recog}
\setlength{\tabcolsep}{4.5pt}
\newcommand*{\rn}[1]{(#1)}
\begin{tabular}{llSSS c SSS c SSS}
    \toprule
    \multicolumn{2}{l}{\mrow{Experiment}} & \multicolumn{3}{c}{WSJ0-2mix} && \multicolumn{3}{c}{WSJ0-3mix} && \multicolumn{3}{c}{WSJ0-4mix} \\
    \cmidrule{3-5}\cmidrule{7-9}\cmidrule{11-13}
    && {CER} & {WER} & {SDRi} && {CER} & {WER} & {SDRi} && {CER} & {WER} & {SDRi} \\
    \midrule
    \rn{1} &DPRNN-TasNet (2 talker) + ASR & 9.6 & 15.8 & 16.5 && {---} & {---} & {---} && {---} & {---} & {---} \\
    \rn{2} &\tbli+ fine-tune ASR & 5.1 & 8.9 & 16.5 && {---} & {---} & {---} && {---} & {---} & {---} \\
    \rn{3} &\tbli+ fine-tune FE + ASR & \thl 4.0 & \thl 7.5 & 11.9 && {---} & {---} & {---} && {---} & {---} & {---} \\ 
    \rn{4} &DPRNN-TasNet (3 talker) + ASR & {---} & {---} & {---} && 31.1 & 49.2 & 10.4 && {---} & {---} & {---} \\
    \rn{5} &\tbli+ fine-tune ASR & {---} & {---} & {---} && 25.5 & 35.3 & 10.4 && {---} & {---} & {---} \\
    \rn{6} &\tbli+ fine-tune FE + ASR & {---} & {---} & {---} && 29.0 & 39.7 & 7.5 && {---} & {---} & {---} \\
    \rn{7} &DPRNN-TasNet (2+3 talker) + ASR & 17.5 & 28.7 & 14.1 && 41.1 & 61.2 & 9.9 && {---} & {---} & {---} \\
    \rn{8} &\tbli+ fine-tune ASR & 7.3 & 11.9 & 14.1 && 19.0 & 27.1 & 9.9 && {---} & {---} & {---} \\
    \rn{9} &\tbli+ fine-tune FE + ASR & 5.8 & 10.5 & 14.4 && 24.9 & 34.9 & 7.8 && {---} & {---} & {---} \\
    \midrule
    \rn{10} &DPRNN-OR-PIT + ASR & 10.8 & 18.3 & \thl 16.7 && 25.2 & 40.7 & \thl 11.9 && 50.5 & 79.9 & \thl 7.8 \\
    % \tbli+ white noise & 5.9 & 11.7 & 16.7 && 15.8 & 28.0 & 11.9 \\
    % \tblii+ fine-tune ASR 2spk & 6.5 & 12.8 & 16.7 && \\
    % \tblii+ fine-tune ASR 2+3spk & \\
    % \tbli+ fine-tune single-iteration ASR 2 spk & 8.0 & 15.7 & 16.7 && 13.4 & 24.5 & 11.9 \\
    % \tblii+ white noise & 7.2 & 14.5 & 16.7 && \\ 
    % \tbli+ fine-tune single-iteration ASR 2+3 spk & 6.5 & 12.7 & 16.7 && 12.2 & 21.2 & 11.9 \\
    % \tblii+ white noise \\
    \rn{11}&\tbli+ VAD & 5.8 & 11.5 & \thl 16.7 && 14.8 & 26.2 & \thl 11.9 && 31.6 & 51.8 & \thl 7.8\\
    \rn{12}&\tblii+ single-iteration fine-tune ASR & 4.3 & 8.7 & \thl 16.7 && \thl 9.3 & \thl 16.3 & \thl 11.9 && 24.6 & 41.1 & \thl 7.8 \\
    \rn{13}&\tblii+ single-iteration fine-tune FE + ASR & 5.2 & 10.0 & 11.9 && 12.2 & 21.2 & 8.9 && 26.2 & 42.6 & 5.1 \\
    \rn{14}&\tblii+ multi-iteration fine-tune ASR & 4.7 & 9.0 & \thl 16.7 && 9.9 & 17.0 & \thl 11.9 && \thl 21.0 & \thl 33.4 & \thl 7.8 \\
    \rn{15}&\tblii+ multi-iteration fine-tune FE + ASR  & 4.6 & 8.7 & 12.9 && 11.8 & 20.5 & 9.0 && 27.8 & 45.2 & 4.8 \\
    \midrule
    \rn{16} &End-to-end ASR \cite{Chang2019_EndtoendMonauralMultispeaker} & {---} & 25.4 & {---} && {---} & {---} & {---} && {---} & {---} & {---} \\
    \rn{17} &Conv-TasNet + ASR \cite{vonNeumann2020_EndtoEndTrainingTime} & 6.0 & 11.1 & 13.8 && {---} & {---} & {---} && {---} & {---} & {---} \\
    \rn{18} &End-to-end ASR \cite{Zhang2020_ImprovingEndtoEndSingleChannel} & {---} & 23.4 & {---} && {---} & {---} & {---} && {---} & {---} & {---} \\
    \midrule
    \rn{19} &Oracle ASR result based on ground truth data & 2.1 & 4.7 & {$+\infty$} && 2.1 & 4.7 & {$+\infty$} && 2.0 & 4.6 & {$+\infty$} \\
    \bottomrule
\end{tabular}
\vspace{-\baselineskip}
\end{table*}

\noindent
The same models are evaluated for counting accuracy in \cref{tab:counting}.
Counting for the 2+3 talker TasNet is performed by an energy-based threshold on its third output.
All thresholds are chosen to maximize the accuracy on \enquote{cv} and \enquote{dev} data.
% \CB{Do you use different thresholds for Counting 1 and 2 Speakers? That may help}
% The OR-PIT is allowed to perform one more iteration than there are speakers in the mixture so that over-estimations are captured correctly.

The \enquote{DPRNN 2+3 talker} model performs well in discriminating between two- and three-talker mixtures, but the threshold does not generalize well to the single-talker scenario due to higher energy levels in the estimated signals.
% There, the energy levels of the \enquote{silent} outputs are larger because the model was not trained on this case.
A counting accuracy of over \SI{90}{\%} is possible with an adjusted threshold, but this degrades the accuracy on two- and three-speaker mixtures.

The OR-PIT detects the number of sources correctly in most cases.
The threshold model performs slightly better for the numbers of talkers it was trained on, but the stop-flag model generalizes better to larger numbers of talkers.
% A threshold often does not generalize to another count of talkers.

Compared to the model selection scheme in \cite{Nachmani2020_VoiceSeparationUnknown}, our system performs better in source counting but worse in separation, where their model achieves \SI{20.1}{dB} SI-SDR improvement and our system \SI{17}{dB} on two talkers.
Similar modifications as \cite{Nachmani2020_VoiceSeparationUnknown} could be applied to the OR-PIT to possibly achieve a similar separation performance.

Although not directly comparable, the stopping criteria introduced in this work seem to perform comparable to the Alexnet classifier with \SI{95.7}{\%} accuracy for less than three talkers \cite{Takahashi2019_RecursiveSpeechSeparation} while being simpler.

\subsection{Single-talker speech recognition}

We use a configuration similar to \cite{Seki2018_PurelyEndtoEndSystem} without the speaker dependent layers for the speech recognizer.
This results in two CNN layers followed by two \gls{BLSTM}P layers with $1024$ units for the encoder, one LSTM layer with 300 units for the decoder and a feature dimension of $80$.
We use a location-aware attention mechanism and ADADELTA \cite{Zeiler2012_ADADELTAadaptivelearning} as optimizer.
Decoding is performed with a word-level RNN language model.
%
% Experiemntal results
%
Our ASR system achieves a WER of \SI{6.4}{\percent} on the WSJ eval92 set.

\subsection{Two-talker speech recognition}

We evaluate the speech recognition performance of the TasNet-based recognizer in rows 1 to 9 of \cref{tab:or-pit-recog}.
% The TasNet and ASR system are pre-trained and then jointly fine-tuned.
We observe similar tendencies as were found in \cite{Heymann2017_BeamnetEndtoendTraining} and \cite{vonNeumann2020_EndtoEndTrainingTime}.
Fine-tuning the ASR part gives a greater improvement than fine-tuning the TasNet but fine-tuning both jointly gives the best performance on two speakers (compare rows 1 to 3) of \SI{58}{\%} relative improvement compared to not fine-tuning.
The final \SI{7.5}{\%} WER (3) forms a new state-of-the-art result on WSJ0-2mix and is a huge improvement over previous techniques \cite{vonNeumann2020_EndtoEndTrainingTime}.
Similar to the separation evaluation, the two-speaker TasNet performs better than the two- and three-speaker model (rows 7 to 9).

\subsection{Multi-talker speech recognition with counting}

After having shown that the joint system performs well with a separator for a fixed number of talkers, we evaluate joint training for the OR-PIT in rows 10 to 15 of \cref{tab:or-pit-recog}.
% The OR-PIT and ASR parts are again separately pre-trained and jointly fine-tuned.
We use the stop-flag OR-PIT due to better counting accuracy on \emph{max} data.

% First, only the ASR part is fine-tuned on two-speaker mixtures or on two- and three-speaker mixtures, where the fine-tuning on two- and three-speaker mixtures improves the performance over using the two-speaker mixtures only.

The OR-PIT shows an odd behavior that is not present in the TasNet.
In regions where only a single talker is active, one output should be silent.
This works for the TasNet, but the OR-PIT outputs the speech signal scaled down heavily.
The ASR system picks it up and creates insertion errors.
An energy-based voice activity detection (VAD) improves the WER substantially to \SI{11.5}{\%} (11) for two talkers without fine-tuning (10).
% A similar effect cannot be observed for the TasNet.

Fine-tuning the ASR system with the single-iteration scheme (12) can improve the performance, but the mismatch created by fine-tuning the FE this way degrades the performance compared to fine-tuning ASR (13).
Using the multi-iteration fine-tuning scheme (14 and 15) can prevent this, but the overall performance is not comparable to the two-speaker TasNet model, although the OR-PIT achieves a better separation performance, according to Tab.~\ref{tab:separation}.
This might be caused by the more complex training procedure.
% There likely is a training procedure or set of hyper-parameters that make the OR-PIT superior to the DPRNN system, but it was not feasible to test more configurations due to limitations in computational resources.

The performance on larger numbers of speakers is listed in the last two columns in \cref{tab:or-pit-recog}.
The OR-PIT again performs better than the TasNet on more than two speakers.
It generalizes to the unseen larger number of four speakers.
Contradicting the results for two speakers, fine-tuning just the ASR part performs best for more than two speakers (rows 5, 8, 12 and 14).
These results suggest that it is beneficial to only fine-tune the ASR back-end in some scenarios, especially when the separation task is more challenging, while fine-tuning the front-end contributes to over-fitting to a specific scenario.

\section{Conclusion}

We build the first joint end-to-end system that performs source number counting and multi-talker speech recognition.
Our specialized model to two talkers provides a new state-of-the-art \gls{WER} for the WJS0-2mix database of \SI{7.5}{\%}.
The OR-PIT-based system that can count the numbers of speakers performs better than the TasNet-based model on three speakers and generalizes well to an unknown number of speakers, i.e., four speakers.
Evaluation of the source counting abilities show very promising performance.

\section{Acknowledgements}
Computational resources were provided by the Paderborn Center for Parallel Computing.

\bibliographystyle{IEEEtran}
\pagebreak  % To prevent balance being called in the second column
\balance
\bibliography{references}

% \begin{thebibliography}{9}
% \bibitem[1]{Davis80-COP}
%   S.\ B.\ Davis and P.\ Mermelstein,
%   ``Comparison of parametric representation for monosyllabic word recognition in continuously spoken sentences,''
%   \textit{IEEE Transactions on Acoustics, Speech and Signal Processing}, vol.~28, no.~4, pp.~357--366, 1980.
% \bibitem[2]{Rabiner89-ATO}
%   L.\ R.\ Rabiner,
%   ``A tutorial on hidden Markov models and selected applications in speech recognition,''
%   \textit{Proceedings of the IEEE}, vol.~77, no.~2, pp.~257-286, 1989.
% \bibitem[3]{Hastie09-TEO}
%   T.\ Hastie, R.\ Tibshirani, and J.\ Friedman,
%   \textit{The Elements of Statistical Learning -- Data Mining, Inference, and Prediction}.
%   New York: Springer, 2009.
% \bibitem[4]{YourName17-XXX}
%   F.\ Lastname1, F.\ Lastname2, and F.\ Lastname3,
%   ``Title of your INTERSPEECH 2020 publication,''
%   in \textit{Interspeech 2020 -- 20\textsuperscript{th} Annual Conference of the International Speech Communication Association, September 15-19, Graz, Austria, Proceedings, Proceedings}, 2020, pp.~100--104.
% \end{thebibliography}

\end{document}